\documentclass[12pt,a4paper,final]{iopart}

\usepackage{iopams}  
\usepackage{graphicx}

\usepackage[colorlinks=true,linkcolor=blue,urlcolor=blue,citecolor=blue]{hyperref}
\usepackage{textcomp}
\usepackage{dsfont}

\begin{document}

\title[Rydberg dressing]{Rydberg dressing: Understanding of collective many-body effects and implications for experiments}

\author{J. B. Balewski, A. T. Krupp, A. Gaj, S. Hofferberth, R. L\"{o}w and T. Pfau}
\ead{j.balewski@physik.uni-stuttgart.de}

\address{
5. Physikalisches Institut, Universit\"{a}t Stuttgart, Pfaffenwaldring 57, 70569 Stuttgart, Germany.}

\begin{abstract}
The strong interaction between Rydberg atoms can be used to control the strength and character of the interatomic interaction in ultracold gases by weakly dressing the atoms with a Rydberg state. Elaborate theoretical proposals for the realization of various complex phases and applications in quantum simulation exist. Also a simple model has been already developed that describes the basic idea of Rydberg dressing in a two-atom basis. However, an experimental realization has been elusive so far. We present a model describing the ground state of a Bose-Einstein condensate dressed with a Rydberg level based on the Rydberg blockade. This approach provides an intuitive understanding of the transition from pure two-body interaction to a regime of collective interactions. Furthermore it enables us to calculate the deformation of a three-dimensional sample under realistic experimental conditions in mean-field approximation. We compare full three-dimensional numerical calculations of the ground state to an analytic expression obtained within Thomas-Fermi approximation. Finally we discuss limitations and problems arising in an experimental realization of Rydberg dressing based on our experimental results. Our work enables the reader to straight forwardly estimate the experimental feasibility of Rydberg dressing in realistic three-dimensional atomic samples.
\end{abstract}

\maketitle

\section{Introduction}
Ultracold atoms provide an ideal system for the simulation of complex many-body systems encountered for example in condensed matter physics \cite{BDN12}. Prime examples are the superfluid to Mott insulator transition \cite{GME02}, studies of the BEC-BCS crossover \cite{NNC10} or the Ising model \cite{SBM11}. So far in such approaches the atoms have mostly been used as hard spheres in various trapping potentials. Long-range interactions \cite{YMG13} have been realized by dipolar atomic species \cite{LKF07} and cold polar molecules \cite{NOM08,ONW10}. A controllable interatomic interaction beyond short-range isotropic character would greatly enrich the available toolbox for quantum simulation and quantum computation. For ground state atoms, the isotropic s-wave interaction potential can be tuned using Feshbach resonances \cite{CGJ10}. Atomic species with a magnetic dipole moment show a combination of an isotropic s-wave interaction potential and a long-range dipolar part \cite{LMS09}. Combining both, the overall character of the interaction can be tuned by changing the strength of the s-wave part \cite{GGP02}. \\
Rydberg atoms in contrast show a very strong van-der-Waals type interaction typically more than ten orders stronger than the interaction between ground state atoms \cite{SWM10}. At short distances or close to a F\"orster resonance there is a transition to a long range dipolar interaction which can exhibit different angular dependencies \cite{CCG04,RLK07}. Furthermore such a F\"orster resonance allows to easily tune the interaction strength \cite{NBK12}. \\
The most obvious problem in using Rydberg excitation in quantum gases is the mismatch in timescales; the lifetime of Rydberg atoms in low l states is on the order of tens of microseconds whereas it typically takes three orders of magnitude longer for a typical experimental system to equilibrate. This problem can be overcome by only weakly dressing \cite{HNP10,PMB10} the atomic ground state with a small fraction $f$ of the Rydberg state enhancing the overall lifetime of the system by $1/f$.
Rydberg dressing has been proposed to realize a number of interesting phases in ultracold gases, such as rotons, solitons or supersolids \cite{HNP10,PMB10,CJB10,MHS11,HCJ12,MGE13,MDL13,MP13}. \\
The strong interaction between Rydberg atoms can result in collective effects in the excitation process, namely the suppression of further excitation in a volume around the first Rydberg atom, the so-called Rydberg-blockade effect \cite{MCT98,TFS04}. The blockade effect has to be included in the calculation of the effective dressing potential as has been already shown by a very simple two-atom model \cite{JR10}. At typical densities in quantum degenerate atomic gases, the number of atoms inside a blockaded volume can become very large. How this affects the dressing potential has been shown by Honer et al. \cite{HWP10}. By numerically solving an effective spin Hamiltonian they found that there is a transition from the pure two-body interaction to a collective N-body regime with the effective interaction vanishing at large densities. \\
In this paper, we develop an analytic N-atom model for the dressing potential that fully accounts for blockade effects. We show that our model gives the same result as the numerical approach in \cite{HWP10}. We use our model to study the effect of Rydberg dressing on a three-dimensional BEC, using realistic experimental parameters. We find that even far from the fully blockaded regime collective effects significantly reduce the interaction induced by Rydberg dressing. This is preventing an observation in current experiments. We provide evidence for this by showing experimental data that also demonstrates further practical problems arising. We discuss the limits of Rydberg dressing in current experiments and point out possible solutions for future approaches.

\section{N-atom model of Rydberg dressing}
Following the argumentation of \cite{JR10}, we begin with the simple case of two atoms, dressed with a Rydberg state by a coupling field with Rabi frequency $\Omega$ and laser detuning $\Delta$. For simplicity we assume the interaction $U(R)$ between atoms in the particular Rydberg state to be of purely repulsive van-der-Waals type. In fact it turns out, that the actual shape of the Rydberg interaction plays a minor role, effectively reduced to the sign of the potential and the value of the blockade radius $r_B$; one can account for attractive interaction by simply changing the sign of the detuning. The Hamiltonian in the dressed states basis $\left|gg\right>$, $1/\sqrt{2}\left(\left|gr\right>+\left|rg\right>\right)$  and $\left|rr\right>$ ($g$: ground state, $r$: Rydberg state) reads:
\begin{equation} \label{eq:2atom}
	H=h\left(
	\begin{array}{ccc}
		0 & \Omega/\sqrt{2} & 0 \\
		\Omega/\sqrt{2} & \Delta & \Omega/\sqrt{2} \\
		0 & \Omega/\sqrt{2} & 2\Delta+U(R)
		\end{array}
	\right)
\end{equation} 
Here the asymmetric singly excited state $1/\sqrt{2}\left(\left|gr\right>-\left|rg\right>\right)$ has been left out, as it is not coupled by $\Omega$. It is straight forward to obtain the new ground state of the system by diagonalizing (\ref{eq:2atom}) depending on the interatomic distance $R$. 
\begin{figure}[tb]
\begin{center}
\includegraphics[width=\textwidth]{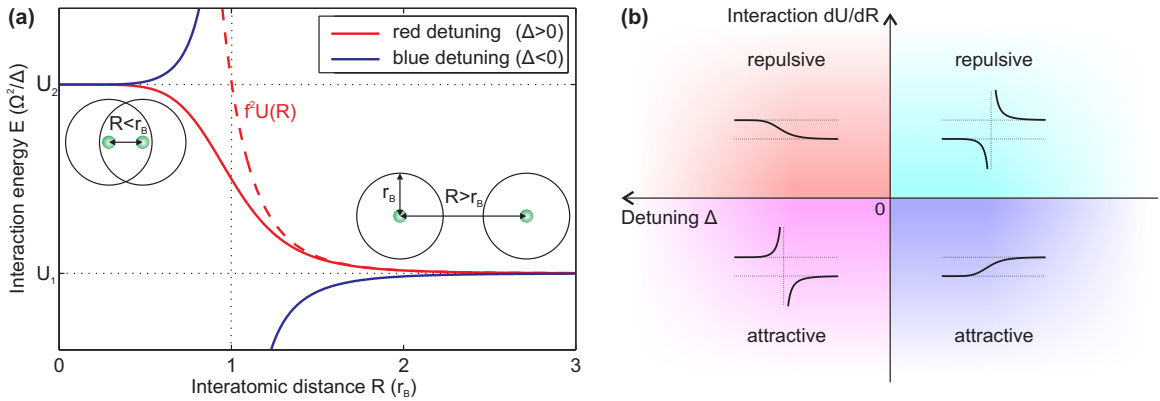}
\caption{\label{fig:2atompot}
Schematic of binary Rydberg dressing potentials. In (a) the interaction potential of a pair of atoms dressed with a repulsive Rydberg state ($\frac{dU}{dR}>0$) by a coupling laser field with Rabi frequency $\Omega$ and detuning $\Delta$ ($\Delta>0$ red, $\Delta<0$ blue). In both cases the interaction potential saturates for very large and very small distances onto a constant value $U_{1/2}$ which is only depending on the laser parameters $\Omega$ and $\Delta$. In case of red detuning and repulsive interaction (or blue detuning and attractive interaction) the potential converges to the Rydberg-Rydberg interaction potential $U(R)$, weighted with the Rydberg fraction $f$ squared (dashed line, corrected for offset $U_1$). Note the scaling of the energy axis with the sign of the detuning $\Delta$. For blue detuning $\Delta<0$ the potential effectively becomes repulsive. The parameter regimes where the four possible potential shapes occur are further illustrated in (b). For red detuning the asymptotic form of the potential is repulsive; for blue detuning it is attractive. Anticrossings appear for the combination of repulsive interaction and blue detuning as well as for attractive interaction and red detuning. 
}
\end{center}
\end{figure}
As can be seen in Figure~\ref{fig:2atompot}a for the combination of blue laser detuning and repulsive potentials this ground state shows a steep anticrossing at around the blockade radius $r_B$ whereas for red detunings there is a smooth step. The blockade radius $r_B$ is defined as the interatomic distance, where the power broadening $\Omega_{\mathrm{eff}}=\sqrt{\Omega^2+(2\Delta)^2}\approx 2|\Delta|$ (for weak dressing $\Omega^2/\Delta^2\ll1$) equates the absolute value of the Rydberg-Rydberg interaction $U(r_B)$. For a repulsive van-der-Waals interaction $U(R)=C_6/R^6$, as it is present for most Rydberg states far from possible resonances, one obtains: 
\begin{equation} \label{rBC6}
	r_B=\left(\frac{C_6}{h\sqrt{\Omega^2+(2\Delta)^2}}\right)^{1/6}\approx\left(\frac{C_6}{h\left|2\Delta\right|}\right)^{1/6}
\end{equation}
In the case of red detuning and repulsive potentials, the potential (up to an offset) is given as the Rydberg interaction potential, weighted with the probability to find both atoms in the Rydberg state, which is at large distances $R$ approximately $f^2=(\Omega^2/(2(\Omega^2+\Delta^2+\Delta\sqrt{\Omega^2+\Delta^2})))^2\approx\frac{\Omega^4}{16\Delta^4}$. At $r_B$ the detuning of the doubly excited state increases due to the Rydberg-Rydberg interaction, tuning this state out of resonance. This is the well known phenomenon of Rydberg blockade \cite{SWM10}. Therefore the probability to find both atoms in the Rydberg state decreases dramatically and the potential saturates at a constant value. For blue detuning the interaction energy saturates on the same asymptotic values, but with the values for small and large interatomic distances $R$ interchanged. It is important to note that the difference between the asymptotic values of the potential energy and therefore also the overall magnitude of the dressing effect is determined only by the laser parameters $\Omega$ and $\Delta$ \cite{BM02,MP13}. These values have been already given by Johnson et al. \cite{JR10}:
	\begin{eqnarray}
		U_1&=h\Delta\left(1-\sqrt{\frac{\Omega^2}{\Delta^2}+1}\right) \label{eq:U2atomfree}\\
		U_2&=h\frac{\Delta}{2}\left(1-\sqrt{\frac{2\Omega^2}{\Delta^2}+1}\right) \label{eq:U2superatom}
	\end{eqnarray}
The overall energy scale is given by the difference of equations (\ref{eq:U2atomfree}) and (\ref{eq:U2superatom}), which can be approximated as $h\frac{\Omega^4}{8\Delta^3}$ for large detunings $\Delta\gg\Omega$. The actual shape and strength of the Rydberg interaction determines mainly the value $r_B$ of the blockade radius and therefore the transition between the two regimes. For attractive Rydberg interaction potentials the same potential shapes appear (see Figure~\ref{fig:2atompot}b). Only the sign of the laser detuning $\Delta$ has to be inverted in Figure~\ref{fig:2atompot}a to account for attractive interaction potentials $U$. \\ 
The two-atom result has a very simple explanation: If the two atoms are separated by a distance much larger than the blockade radius $r_B$, the system can be described as two independent atoms in a laser field with Rabi frequency $\Omega$ and detuning $\Delta$ with respect to the transition to the Rydberg level. Therefore both atoms individually exhibit an AC Stark shift $U_1/2$. If the two atoms approach each other, at the distance of the blockade radius $r_B$ (equation~\ref{rBC6}) the Rydberg-Rydberg interaction potential becomes so strong that the state with both atoms excited is tuned out of resonance. The explanation of the behaviour of the system in the blockaded regime becomes obvious from the following discussion of the $N$-atomic case. \\
We study now the density dependent energy of a system consisting of $N$ Rydberg dressed atoms. At very low densities the interatomic distances $R$ are much larger than the blockade radius $r_B$. The atoms are therefore independent and the energy of the system is $N$ times the light shift $U_1/2$ of one single atom in equation (\ref{eq:U2atomfree}):
\begin{equation} \label{eq:UNsuperatomfree}
	U_1(N)=Nh\frac{\Delta}{2}\left(1-\sqrt{\frac{\Omega^2}{\Delta^2}+1}\right)
\end{equation}
In the high density limit all atoms are situated within one blockade sphere of volume $\frac{4}{3}\pi r_B^3$. In this case, the dimensionality of the Hilbert space can be dramatically reduced by the fact that all states with more than one Rydberg excitation are completely tuned out of resonance due to the Rydberg blockade. The resulting Hamiltonian in the $(N+1)$-dimensional basis $\left|gg...g\right>$, $\left|rg...g\right>$, $\left|gr...g\right>$,... , $\left|gg...r\right>$ reads:
\begin{equation}
	H_N=h\left( \begin{array}{ccccc}
	0 & \frac{\Omega}{2} & \frac{\Omega}{2} & \ldots &  \frac{\Omega}{2} \\
	\frac{\Omega}{2} & \Delta & 0 & \ldots & 0 \\
	\frac{\Omega}{2} & 0 & \ddots & \ddots & \vdots \\
	\vdots & \vdots & \ddots &\Delta & 0 \\
	\frac{\Omega}{2} & 0 &\ldots &  0 &\Delta 
	\end{array} \right)
\end{equation}
It can be shown, e.g. by mathematical induction, that the polynomial determining the eigenenergies E is given as:
\begin{equation}
	\det\left(H_N-\mathds{1}E\right)=\left[E^2-h\Delta E-h^2\frac{N\Omega^2}{4}\right]\left(h\Delta-E\right)^{N-1}
\end{equation}
The energy of the ground state of the fully blockaded N-atomic state therefore is
\begin{equation}
	U_2(N)=h\frac{\Delta}{2}\left(1-\sqrt{\frac{N\Omega^2}{\Delta^2}+1}\right)
	\label{eq:UNsuperatom}
\end{equation}
Comparing this result to equations~(\ref{eq:U2atomfree}) and (\ref{eq:U2superatom}), the explanation of the energy in the limit of full blockade becomes obvious. Due to the Rydberg blockade all N atoms share one Rydberg excitation forming a collective state $\frac{1}{\sqrt{N}}\sum_{i=1}^{N}\left|g_1,g_2,...,r_i,...,g_{N}\right>$. This state, a so called super atom \cite{Vul06}, is coupled by the light field to the ground state with a collectively enhanced Rabi frequency $\sqrt{N}\Omega$. The collective Rabi frequency then also has to be used in the calculation of the AC Stark effect. The energy of the fully blockaded N-atomic system (equation~\ref{eq:UNsuperatom}) is therefore the light shift of a single super atom with a collective Rabi frequency $\sqrt{N}\Omega$ and detuning $\Delta$. In the special case $N=2$ the result of equation~(\ref{eq:U2superatom}) is recovered. The interaction potential induced by Rydberg dressing shown in Figure~\ref{fig:2atompot}a can therefore be viewed as the gradual transition between a collective light shift of one super atom and the individual light shift of two independent atoms in the vicinity of the blockade radius $r_B$. \\ 
To study the consequences of high densities for the effect of Rydberg dressing, we calculate now the additional interaction energy $E_{\mathrm{dress}}$ per atom. The number of atoms $N$ within one blockade sphere of radius $r_B$ is determined by the ground state atom density $\rho$:
\begin{equation} \label{eq:Nrho}
	N=\frac{4}{3}\pi r_B^3 \rho=\frac{\rho}{f\rho_B}
\end{equation}
where $\rho_B$ denotes the critical density, where according to \cite{HWP10} blockade phenomena start to play a role. This is when the average number of Rydberg atoms in a blockade sphere $f\cdot \rho\cdot4/3\pi r_B^3$ approaches unity:
\begin{equation} \label{eq:rhoB}
	\rho_{B}=\frac{3}{\pi r_B^3}\frac{\Delta^2}{\Omega^2}=\frac{3\sqrt{2h}}{\pi\sqrt{C_6}}\frac{|\Delta|^{5/2}}{\Omega^2}
\end{equation}        
The difference of light shift between one super atom $U_2(N)$ (equation \ref{eq:UNsuperatom}) and N independent atoms $U_1(N)$ (equation \ref{eq:UNsuperatomfree}) is shared among all $N$ atoms, so that the contribution per atom reads:
\begin{equation} \label{eq:EdressN}
	E_{\mathrm{dress}}(\rho)=  h\frac{\Delta}{2}\left(\frac{\Omega^2}{4\Delta^2}\frac{\rho_B}{\rho}\left(1-\sqrt{4\frac{\rho}{\rho_B}+1}\right)-\left(1-\sqrt{\frac{\Omega^2}{\Delta^2}+1}\right)\right) 
\end{equation}
Note that this expression is only valid for $\rho/\rho_B\geq\Omega^2/(4\Delta^2)$; at lower densities the blockade does not play a role any more and the mean interaction energy per particle is given as $f^2U(\rho^{-1/3})\ll1$.
The relevant quantity for calculating the impact on an ultracold sample (see section~\ref{effBEC}) is the variation of the energy density $E_{\mathrm{eff}}(\rho)=\rho E_{\mathrm{dress}}(\rho)$ with density:
\begin{equation} \label{eq:Eeff}
	\partial_{\rho}E_{\mathrm{eff}}(\rho)=h\frac{\Delta}{2}\left(\sqrt{\frac{\Omega^2}{\Delta^2}+1}-1 -\frac{\Omega^2}{2\Delta^2}\frac{1}{\sqrt{4\frac{\rho}{\rho_B}+1}}\right)
\end{equation}
\begin{figure}[tb]
\begin{center}
\includegraphics[width=0.6\textwidth]{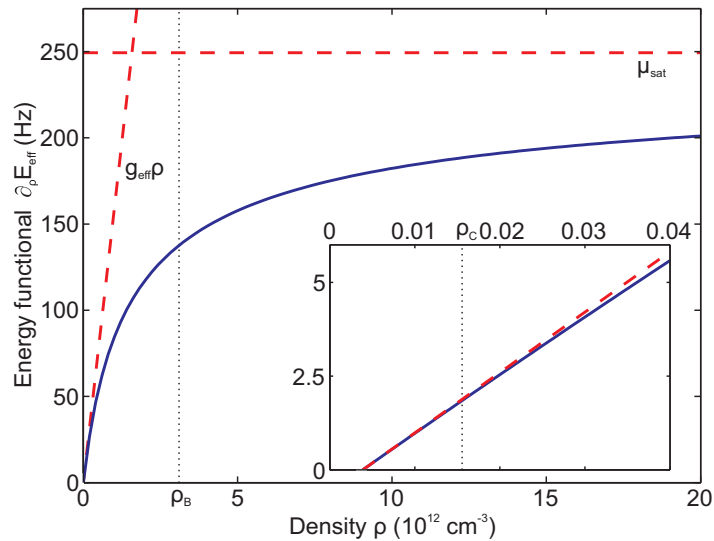}
\caption{\label{fig:Eeff}
Rydberg dressing induced energy functional $\partial_{\rho}E_{\mathrm{eff}}(\rho)$ versus ground state atom density $\rho$. The dashed lines show the asymptotic behaviour at high and low densities. The inset shows a zoom at very low densities. The values are calculated for a Rabi frequency of $\Omega=10\,\mathrm{kHz}$, a red detuning of $\Delta=100\,\mathrm{kHz}$ and a purely repulsive van-der-Waals interaction with $C_6=1.89\cdot10^{-28}\,\mathrm{Hzm}^6$, corresponding to the 35S Rydberg state \cite{SSW05} and resulting in a blockade radius of $r_B=3.1\,\textrm{\textmu m}$. The critical densities $\rho_B$ and $\rho_C$ are explained in the text. 
}
\end{center}
\end{figure}
The density dependence of this quantity is shown in Figure~\ref{fig:Eeff}. For low densities $\rho$ and weak dressing $\Omega^2/\Delta^2\ll1$, there is a linear increase with slope
\begin{equation} \label{gdress}
	g_{\mathrm{eff}}=h\frac{\pi}{6}r_B^3\frac{\Omega^4}{\Delta^3}
\end{equation} 
This slope deviates from the value in \cite{HWP10} only by a constant factor of $\pi/2$. At higher densities, the last part of equation~(\ref{eq:Eeff}) vanishes and the energy functional $\partial_{\rho}E_{\mathrm{eff}}$ quickly saturates on a constant value 
\begin{equation} \label{eq:Edresslim}
	\mu_{\mathrm{sat}}=h\frac{\Delta}{2}\left(\sqrt{\frac{\Omega^2}{\Delta^2}+1}-1\right)\approx h\frac{\Omega^4}{2\Delta^3}
\end{equation}
This expression agrees with the result in ~\cite{HWP10}. The energy $E_{\mathrm{dress}}(\rho)$ from equation~(\ref{eq:EdressN}) saturates also exactly on the same value. This is the AC Stark shift of a single free Rydberg dressed atom, as the contribution of the dressed super atom to the energy of a single atom becomes negligible. \\       
As can be seen in Figure~\ref{fig:Eeff} the energy functional $\partial_{\rho}E_{\mathrm{eff}}$ deviates from the initial linear slope already at densities well below $\rho_{\mathrm{B}}$. The simple explanation is that collective effects already start to play a role as soon as there are more than two atoms within one blockade volume, which is the case at densities above
\begin{equation}
	\rho_{C}=\frac{3}{2\pi r_B^3} =\frac{3}{2\pi} \sqrt{\frac{2|\Delta|}{C_6}}
\end{equation}
As a consequence, collective effects beyond two-body interaction have to be considered also at relatively low densities, where they already start to reduce any influence of Rydberg dressing.

\section{Rydberg dressing of Bose-Einstein condensates} \label{effBEC}
We now apply our results to calculate the modification of the wavefunction of a Bose-Einstein condensate under Rydberg-dressing. Our model provides an analytic expression of the steady state density distribution of a BEC dressed with a homogeneous coupling field $\Omega$ even in case of a cylindrically symmetric harmonic trapping potential. This not only allows us to predict effects in realistic experimental situations but also to extract the scaling of the deformation with different parameters. \\
We consider a BEC in a three dimensional harmonic trap. The evolution of the condensate wavefunction $\psi(\vec{r},t)$ is described by the Gross-Pitaevskii equation:
\begin{equation} \label{eq:GPEt}
	i\hbar\frac{\partial\psi}{\partial t}= \left[-\frac{h^2\nabla^2}{2m}+ \sum_{i=1}^3\frac{m\omega_i^2x_i^2}{2}+g_s\left|\psi\right|^2+\partial_{\rho}E_{\mathrm{eff}}(\left|\psi\right|^2)\right]\psi
\end{equation}
The first two terms are the usual kinetic energy contribution and the harmonic trapping potential, characterized by the three oscillation frequencies $\omega_i$. The third term is the mean-field contribution of the contact interaction between the atoms, while the fourth term $\partial_{\rho}E_{\mathrm{eff}}(\rho)$ from equation~(\ref{eq:Eeff}) describes the effect of Rydberg dressing on the BEC. Besides mean-field approximation this incorporates two further approximations: First of all, local density approximation requires the sample to be much larger than the length scale of the interaction given by the blockade radius $r_B$ (equation~\ref{rBC6}). Secondly the actual shape of the Rydberg interaction potential is neglected by assuming a step between the two asymptotic values of the dressing potential at the blockade radius $r_B$. \\
Even without solving the equation some important conclusions can be drawn. In the very low density regime the interaction energy is linear in the density $|\psi|^2$ and can therefore be described with an effective s-wave scattering length $\frac{m g_{\mathrm{eff}}}{4\pi \hbar^2}$ (see equation~\ref{gdress}). For high densities, the energy functional $\partial_{\rho}E_{\mathrm{eff}}(\rho)$ becomes constant and thus only contributes as a constant offset to the chemical potential $\mu$. In this case no effect of Rydberg dressing on the density distribution can be expected. \\
The ground state stationary solution $\psi(\vec{r},t)=\psi(\vec{r}) e^{-i\mu/\hbar t}$ of equation~\ref{eq:GPEt} can be obtained numerically e.g. using the split step Fourier method \cite{J96}. For condensates with large atom numbers the trapping potential and the interaction energy are large \cite{KDS99}. 
\begin{figure}[tb]
\begin{center}
\includegraphics[width=\textwidth]{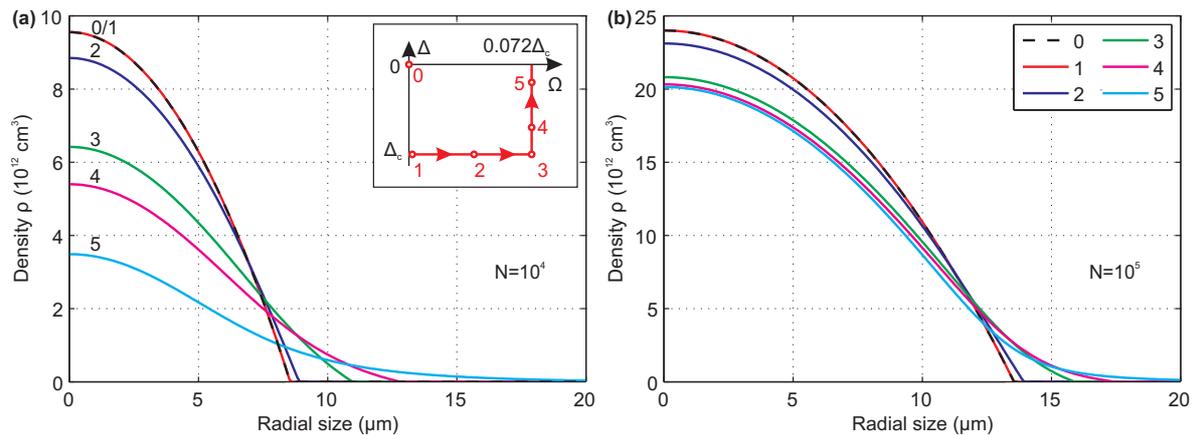} 
\caption{\label{fig:Honer}
Reproduction of Figure~3 of \cite{HWP10} with the super atom model: Density profile $\rho(r)$ of a Bose-Einstein condensate consisting of $N=10^4$ (a) and $N=10^5$ (b) atoms \cite{error} in a radially symmetric trap with trap frequency $\omega=2\pi\cdot15.9\,\mathrm{Hz}$. The inset in (a) shows the parameters of the laser field coupling to the Rydberg level in the plane of Rabi frequency $\Omega$ and detuning $\Delta$ in units of the critical detuning $\Delta_c=107\,\mathrm{kHz}$. For larger condensates and therefore higher densities $\rho$ the effect of Rydberg dressing is strongly reduced.
}
\end{center}
\end{figure} 
In Thomas-Fermi approximation then the kinetic energy term can be neglected which is usually a good approximation except for the low density wings of the condensate \cite{VBL98}.  
Without the additional energy functional $\partial_{\rho}E_{\mathrm{eff}}$ the equation becomes linear in $|\psi|^2$ making the analytic solution particularly simple. The well known result is a paraboloidal density distribution $\rho=|\psi|^2$ where the value of the chemical potential $\mu$ is determined by the normalization to the total atom number $N$. With the contribution from Rydberg dressing (equation~\ref{eq:Eeff}) the calculation of the modified ground state density $\rho$ becomes only slightly more involved. For red detunings $\Delta>0$ equation~(\ref{eq:GPEt}) can be rewritten as a cubic equation in the density $\rho$ that can be efficiently solved analytically. Only the chemical potential $\mu$ has to be calculated numerically under the constraint that the total BEC atom number remains constant.
In Figure~\ref{fig:Honer}, we show the resulting condensate density distributions when using the same parameters as in \cite{HWP10}. Our analytic model is in very good agreement with the numerical results by Honer et al. also obtained within Thomas-Fermi approximation, but there is a finite deviation. Nevertheless our results should be more than enough precise to allow predictions about overall scaling and orders of magnitude. However, since the additional term $\partial_{\rho}E_{\mathrm{eff}}$ of Rydberg dressing is expected to be small, it is not obvious that Thomas-Fermi approximation is valid here. 
\begin{figure}[tb]
\begin{center}
\includegraphics[width=\textwidth]{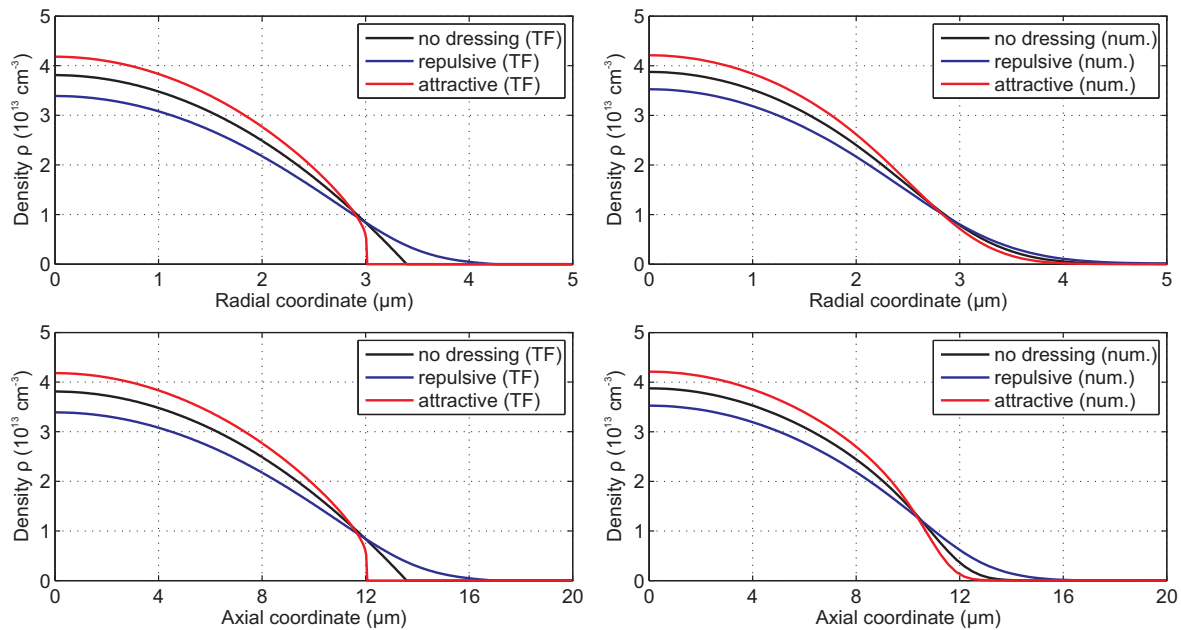}
\caption{\label{fig:dressvgl}
Radial and axial density profiles of a BEC consisting of $10^4$ $^{87}\mathrm{Rb}$ atoms in an axially symmetric trap with trap frequencies $\omega_r=2\pi\cdot 80\,\mathrm{Hz}$ and $\omega_z=2\pi\cdot 20\,\mathrm{Hz}$. The analytic result using Thomas-Fermi approximation (left) is compared to the full numerical solution of the Gross-Pitaevskii equation in three dimensions (right). The unperturbed density distribution as well as two different Rydberg dressed situations are shown. The repulsive case has been calculated for the same parameters as Figure~\ref{fig:Eeff} (35S, $\Omega=10\,\mathrm{kHz}$, $\Delta=100\,\mathrm{kHz}$). For the attractive case the sign of the detuning has been inverted.
}
\end{center}
\end{figure}
As an exemplary check, we calculate the density distribution for typical parameters using both methods as shown side by side in Figure~\ref{fig:dressvgl}. As expected a repulsive dressing potential leads to an expansion of the condensate whereas an attractive potential makes the cloud become smaller and denser. Furthermore the full numerical solution and the result of Thomas-Fermi approximation agree very well in the centre of the condensate, whereas the deviation at the outer parts becomes significant. It turns out that in these regions the kinetic energy term is dominating the density distribution so that here the effect of Rydberg dressing is even less visible than in the centre of the BEC. This is in contrast to what one could have expected from the scaling of the energy functional $\partial_{\rho}E_{\mathrm{eff}}(\rho)$ with density alone (see Figure~\ref{fig:Eeff}). However, the fact that the impact of Rydberg dressing is vanishing at higher densities can be seen comparing the attractive and repulsive dressing. The effect of repulsive Rydberg dressing is slightly stronger since it becomes self-amplifying. \\
Since the modification of the density distribution by the Rydberg dressing is the strongest in the centre of the condensate we use in the following the relative change of peak density to quantify the response of a three-dimensional asymmetric BEC. As the behaviour of this quantity is well reproduced using Thomas-Fermi approximation, we can make use of the significant speed up provided by the analytic calculation within Thomas-Fermi approximation. Furthermore, the residual deviation from the more exact numerical simulation further decreases at higher atom numbers in the condensate. 

\section{Systematic study of parameter space} \label{discussion}
We can now compute the deformation of a BEC depending on different experimentally accessible parameters. These are essentially the laser parameters, the Rabi frequency $\Omega$ and detuning to the Rydberg state $\Delta$, the initial peak density $\rho_0$ of the condensate and the blockade radius $r_B$ which within the model of equation~(\ref{rBC6}) can be controlled via the $C_6$ coefficient of the Rydberg state. \\
From Figure~\ref{fig:Eeff} one can conclude that at high atomic densities the effect of Rydberg dressing is vanishing. The Rydberg blockade here limits the achievable interaction strength. In a not fully blockaded sample, at densities below $\rho_B$, a higher density increases the collective Rabi frequency $\sqrt{N}\Omega$ leading to more Rydberg excitation in the system. Above $\rho_B$, however, no further Rydberg excitation is possible. The total interaction energy, determined by the number of Rydberg excitations in the system, is then saturated on a constant value that is shared among more atoms.  The contribution for each individual atom is thus effectively vanishing at high densities. At very low densities, however, the interatomic distance increases thereby also reducing the interaction strength. The maximum effect is therefore expected in the intermediate regime where the energy difference per atom between the blockaded system and the non blockaded system is the highest. We estimate this region to be at the point where the system just starts to become fully blockaded, at densities around $\rho=\rho_B$. \\
The laser parameters $\Omega$ and $\Delta$ mainly determine the fraction of Rydberg excitations $f\approx\Omega^2/4\Delta^2$. Large Rabi frequencies $\Omega$ and small detunings $\Delta$ to the Rydberg state are therefore increasing the effect of Rydberg dressing. However, there is an upper bound of the tolerable Rydberg fraction $f$ given by the decay of the Rydberg state.  Large Rydberg fractions $f$ lead to a strongly reduced lifetime $\propto 1/f$ of the dressed state that prevents the observation of any mechanical effect. This constraint is further discussed in part~\ref{problems}.\\ 
In order to quantify the effect of Rydberg dressing onto a BEC we calculate the relative change $\Delta\rho/\rho_0$ of the peak density of the condensate. 
\begin{figure}[tb]
\begin{center}
\includegraphics[width=0.6\textwidth]{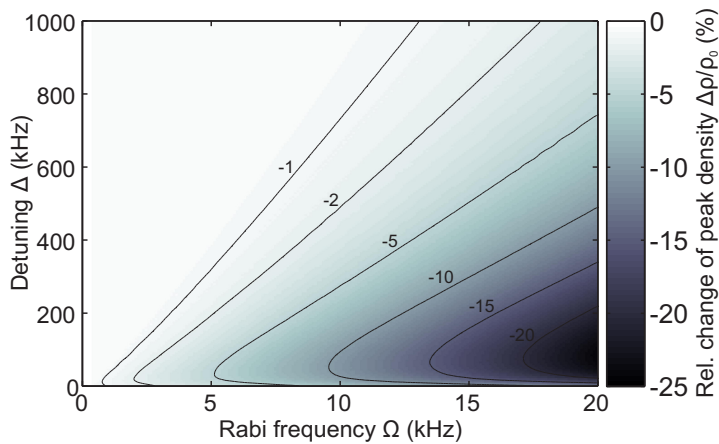}
\caption{\label{fig:depOmegaDelta}
Dependency of Rydberg dressing of a BEC on laser parameters $\Omega$ and $\Delta$. The relative change of peak density $\Delta\rho/\rho_0$ in steady state is calculated for a condensate consisting of $N=2\cdot10^4$ atoms in a cylindrically symmetric trap with trap frequencies $\omega_r=2\pi\cdot 20\,\mathrm{Hz}$ and $\omega_z=2\pi\cdot 80\,\mathrm{Hz}$. The condensate is dressed with a repulsive Rydberg state with $C_6=6.10\cdot10^{-29}\,\mathrm{Hz}\mathrm{m}^6$, corresponding to the 32S Rydberg state. 
}
\end{center}
\end{figure}
The dependency of this quantity on the Rabi frequency $\Omega$ and the laser detuning $\Delta$ to the Rydberg state are shown in Figure~\ref{fig:depOmegaDelta}. The range of Rabi frequencies is given by technical constraints as discussed in part~\ref{experiment}. As expected, the effect increases towards higher Rabi frequencies $\Omega$ and lower detunings $\Delta$. At very small detunings $\Delta\leq\Omega$, however, there is some deviation from this trend since here the blockade radius $r_B$ according to equation~(\ref{rBC6}) becomes large. Anyway in this parameter region the approximation of weak dressing $f\ll 1$ is not fulfilled any more. \\
\begin{figure}[tb]
\begin{center}
\includegraphics[width=\textwidth]{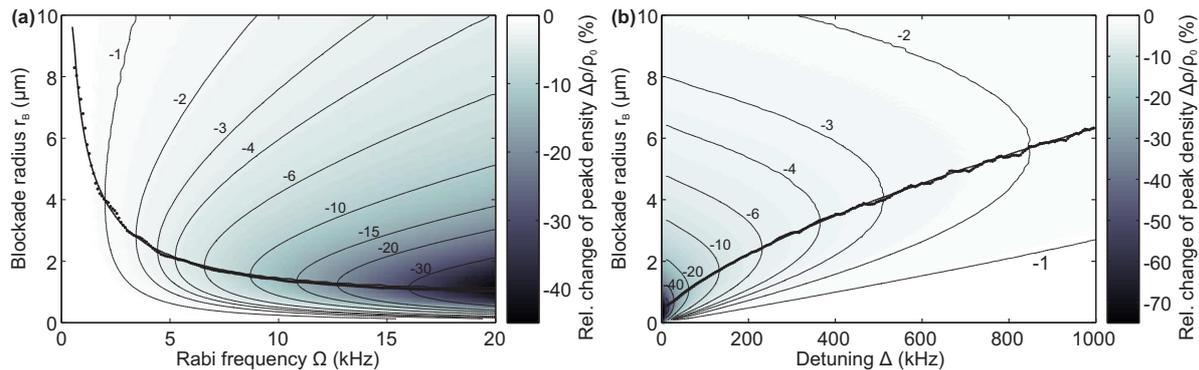}
\caption{\label{fig:depOmegaDeltarB}
Dependency of Rydberg dressing of a BEC on the Rabi frequency $\Omega$ (a), laser detuning $\Delta$ (b) and the blockade radius $r_B$ as an independent quantity. The relative change of peak density $\Delta\rho/\rho_0$ in steady state is calculated for a condensate consisting of $N=2\cdot10^4$ atoms in a cylindrically symmetric trap with trap frequencies $\omega_r=2\pi\cdot 20\,\mathrm{Hz}$ and $\omega_z=2\pi\cdot 80\,\mathrm{Hz}$. The condensate is dressed detuned by $\Delta=100\,\mathrm{kHz}$ (a) with Rabi frequency $\Omega=10\,\mathrm{kHz}$ (b) to a repulsive Rydberg state with $C_6=6.10\cdot10^{-29}\,\mathrm{Hz}\mathrm{m}^6$, corresponding to the 32S Rydberg state. The black dots show the blockade radii with maximum effect at fixed $\Omega$ and $\Delta$ respectively. The black solid line shows the value of $r_{B,m}$ (equation~\ref{eq:rBm}) when the sample on average starts to become fully blockaded. This is where the maximum effect is expected (see text). 
}
\end{center}
\end{figure} 
So far we assumed that the finite blockade radius $r_B$ is given by power broadening according to equation~(\ref{rBC6}) neglecting technical sources of laser broadening. In Figure~\ref{fig:depOmegaDeltarB} we study also the influence of the blockade radius $r_B$ as an independent quantity. The effect increases monotonously with Rabi frequency $\Omega$ and decreases with detuning $\Delta$. For each value of $\Omega$ and $\Delta$, however, there is a clear maximum in the blockade radius $r_B$. This can be explained by the fact that the blockade radius $r_B$ determines the length scale of the system. The density $\rho$ is rescaled by the blockade radius $r_B$ according to equation~(\ref{eq:Nrho}). As discussed before, the maximum effect of Rydberg dressing is expected when the system becomes fully blockaded at $\rho=\rho_B$. Keeping the density $\rho$ fixed we can calculate the optimal blockade radius $r_{B,m}$ from equation~(\ref{eq:rhoB}):
\begin{equation} \label{eq:rBm}
	r_{B,m}=\sqrt[3]{\frac{3}{\pi\rho}\frac{\Delta^2}{\Omega^2}}
\end{equation}
It turns out that this even quantitatively reproduces the dependency observed in Figure~\ref{fig:depOmegaDeltarB} if we use $3/4$ times the mean density of the dressed condensate. For small deformation of the condensate also the mean density $\overline{\rho}=2/5\rho_0$ of the initial Thomas-Fermi density distribution with peak density $\rho_0$ is a good approximation.

\section{Dipolar interaction} \label{Dipolar}
\begin{figure}[tb]
\begin{center}
\includegraphics[width=0.6\textwidth]{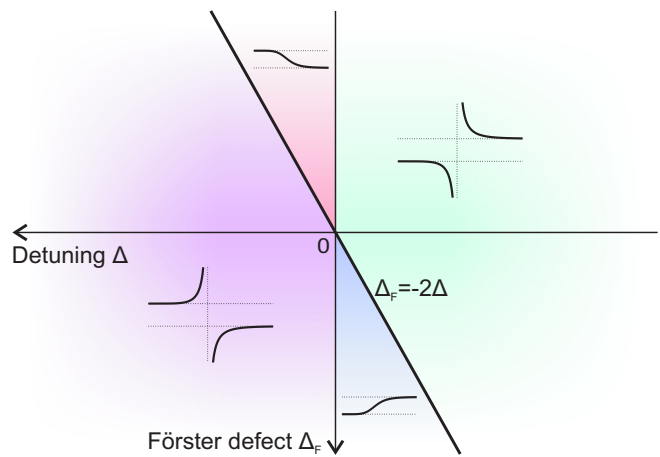}
\caption{\label{fig:Foersterscheme}
Scheme of different potentials of a pair of atoms dressed with a Rydberg state close to a F\"orster resonance. In the plane of detuning $\Delta$ and F\"orster defect $\Delta_F$ four different regimes are identified. The regime of a smooth potential is narrowed down compared to Figure~\ref{fig:2atompot}b to small absolute detunings $|\Delta|<|\Delta_F|/2$. }
\end{center}
\end{figure}
Within the model presented here, the shape of the interaction potential only determines the value of the blockade radius $r_B$ and the sign of the Rydberg dressing. Therefore one would not expect that a Rydberg-Rydberg interaction of dipolar type could lead to fundamentally new effects. However, it seems worthwhile to consider the dipolar interaction of Rydberg states close to a F\"orster resonance. Some Rydberg states can be tuned into resonance by applying microwaves \cite{BPM07,TDN08} or small electric fields \cite{ARM02,ABV04,ABP06,CSN06,RTB10,RYL08}. First of all, the interaction between two Rydberg atoms close to such a resonance shows an angular dependency \cite{CCG04,RLK07} that is expected to translate directly into the Rydberg dressed interaction. Secondly and more importantly the strength and sign of the interaction close to the F\"orster resonance can be tuned via the F\"orster defect $\Delta_F$ \cite{NBK12}. This would not only allow to easily realize repulsive and attractive Rydberg dressing while keeping other parameters of the system like e.g. the Rabi frequency $\Omega$ constant, but also enable to control the blockade radius $r_B$ of the system by changing the interaction strength. As the discussion above showed, this is a key parameter that rescales the important density scale of the problem. \\
Yet in case of a F\"orster resonance the level scheme becomes slightly more involved and the two-atom Hamiltonian (\ref{eq:2atom}) has to be extended by the pair state $\left|r'r''\right>$ which on F\"orster resonance ($\Delta_F=0$) becomes degenerate with the doubly excited Rydberg state $\left|rr\right>$:
\begin{equation} \label{eq:2atomFoerster}
	H=h\left(
	\begin{array}{cccc}
		0 & \Omega/\sqrt{2} & 0 & 0\\
		\Omega/\sqrt{2} & \Delta & \Omega/\sqrt{2} & 0\\
		0 & \Omega/\sqrt{2} & 2\Delta & U_{dd}(R) \\
		0 & 0 & U_{dd}(R) & 2\Delta+\Delta_F 
		\end{array}
	\right)
\end{equation}
where $U_{dd}(R)=C_3/R^3$ is the coupling of the two doubly excited Rydberg states. Negative F\"orster defects $\Delta_F<0$ lead to repulsive interaction, attractive interaction can be realized by choosing positive F\"orster defects. It turns out that similar potential curves as in Figure~\ref{fig:2atompot}a can be obtained. In particular the asymptotic values are the same as for the Hamiltonian~(\ref{eq:2atom}). However, the smooth dressing potential (red curve in Figure~\ref{fig:2atompot}a) is only present in the regime of small absolute detunings $|\Delta|<|\Delta_F|/2$. For larger detunings $\Delta$ there is an additional anticrossing occurring. The regimes of different two body potentials in the plane of detuning $\Delta$ and F\"orster defect $\Delta_F$ are sketched in Figure~\ref{fig:Foersterscheme}. However, we note that the anticrossing is likely to play a role only in very cold atomic samples since fast atoms follow the potential curves diabatically \cite{W05}. Then again only the asymptotic values of the potential and the blockade radius $r_B$ are important. In this case, the sign of the Rydberg dressed potential can be fully controlled with the sign of the laser detuning $\Delta$ alone, also in the absence of a F\"orster resonance.

\section{Current experimental situation} \label{experiment}
In our present experimental apparatus described in \cite{LWN12}, we can realize Bose-Einstein condensates in a cylindrically symmetric trap with a radial trapping frequency $\omega_r=2\pi\cdot 22\,\mathrm{Hz}$ and axial trapping frequencies down to $\omega_z=2\pi\cdot 82\,\mathrm{Hz}$. Typical atom numbers vary around $10^5$ corresponding to a peak density of $10^{14}\,\mathrm{cm}^{-3}$. Atom numbers down to $2\cdot10^4$ can be realized in order to reduce the peak density to $5\cdot10^{13}\,\mathrm{cm}^{-3}$ at the expense of increasing atom number fluctuations observed. We excite Rydberg S and D-states via a two-photon transition detuned by $500\,\mathrm{MHz}$ from the intermediate $5P_{3/2}$ state using continuous wave diode lasers at wavelengths of $780\,\mathrm{nm}$ and $480\,\mathrm{nm}$. Depending on the Rydberg state, the Rabi frequency of the upper transition is limited by technical constraints to several tens of MHz. The focused $480\,\mathrm{nm}$ laser is switched on adiabatically during $400\,\mathrm{ms}$ since it creates a dipole potential for the ground state atoms \cite{SWM10}. As it is common for all Rydberg excitation schemes based on a two-photon transition \cite{LWN12,VRB11}, the Rabi frequency of the lower transition is limited by off-resonant scattering from the intermediate state \cite{GWO00} as in our case, or two-photon ionization \cite{ACC04}. The first effect can be estimated if reabsorption of the emitted photons \cite{CCL98} is taken into account. In the experiment we choose the red laser power such that the total atom loss over the length of the laser pulse is largely negligible. For experiments lasting $100\,\mathrm{ms}$ we can apply effective Rabi frequencies on the order of few kHz. This sequence length corresponds to about twice the inverse axial trap frequency and was chosen to allow for the BEC density distribution to reach its equilibrium. We switch off the Rydberg lasers and the trap simultaneously. After a time of flight lasting $50\,\mathrm{ms}$, an absorption image is taken. From these images we extract the change of the BEC atom number and the aspect ratio similar to the procedure in \cite{BKG13} in order to compare them to reference measurements with the $780\,\mathrm{nm}$ laser off. \\
\begin{figure}[tb]
\begin{center}
\includegraphics[width=1\textwidth]{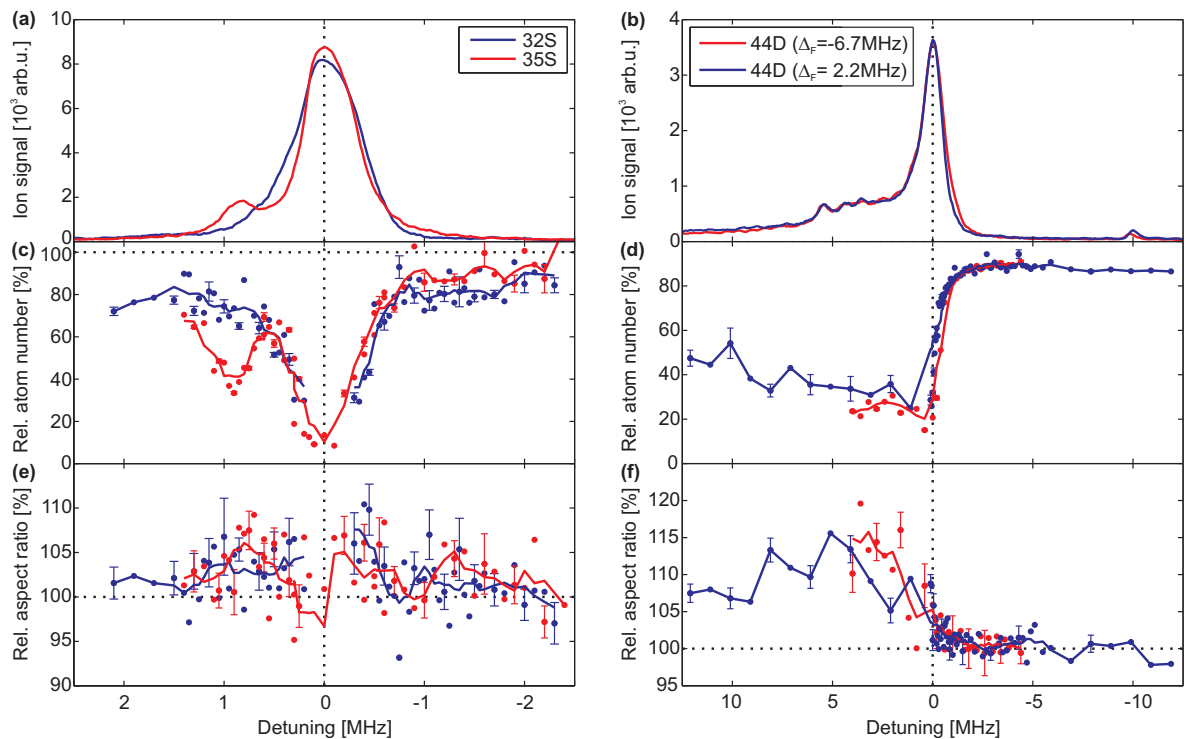}
\caption{	\label{fig:Dressingall} 
	Experimental results on dressing different Rydberg states to a Bose-Einstein condensate: The condensate consisting of $2\cdot10^4$ (S-states, left panels) and $8\cdot10^4$ (D-state, right panels) atoms is dressed for $100\,\mathrm{ms}$ at fixed Rabi frequencies (32S$_{1/2},\,m_J=1/2$: $\Omega=3.4\,\mathrm{kHz}$, 35S$_{1/2},\,m_J=1/2$: $\Omega=2.3\,\mathrm{kHz}$, 44D$_{5/2},\,m_J=5/2$: $\Omega=2.6\,\mathrm{kHz}$) and variable detuning $\Delta$ to the Rydberg state respectively. The relative change of BEC atom number (c, d) and aspect ratio (radial/axial width, e, f) was extracted from absorption images taken after a time of flight of $50\,\mathrm{ms}$. In (a, b) the ion signal from reference measurements in a thermal sample is shown. The parameters were chosen to make tiny signals from molecular states visible (S-states (a): excitation pulse length $100\,\textrm{\textmu s}$, Rabi frequencies $\Omega=17.5\,\mathrm{kHz}/2.3\,\mathrm{kHz}$; D-state (b): excitation pulse length $5\,\textrm{\textmu s}$, Rabi frequency $\Omega=8.5\,\mathrm{kHz}$). Therefore the signal on resonance is highly saturated. The solid lines in (c-f) are a moving average proportional to the excitation linewidth (width $0.3\,\mathrm{MHz}$, c, e; $1\,\mathrm{MHz}$, d, f) as guide to the eye. Error bars, $\pm1$ s.d. from eight independent measurements.
}
\end{center}
\end{figure}
The results for two different Rydberg S-states and one Rydberg D-state close to a F\"orster resonance are shown in Figure~\ref{fig:Dressingall} in comparison to reference Rydberg spectra measured by field ionization and ion detection in a thermal sample as described in \cite{BKG13,LWN12}. For the S-states and the Rabi frequencies given in Figure~\ref{fig:Dressingall}, the onset of full blockade according to equations~(\ref{rBC6}) and (\ref{eq:rBm}) using the mean density $\overline{\rho}=2/5\rho_0$ is taking place at detunings $\Delta$ just below $100\,\mathrm{kHz}$. This is the regime where the largest effect can be expected (see part~\ref{discussion}). Here obviously the decay from the Rydberg state, visible as a loss feature in Figure~\ref{fig:Dressingall}c, is the limiting factor; the decay leads to strong heating that destroys nearly the whole condensate at small detunings $|\Delta|<200\,\mathrm{kHz}$. Outside this regime the expected effect of Rydberg dressing according to Figure~\ref{fig:depOmegaDelta} is very small. Consequently we are not able to detect any  significant deformation of the condensate within the experimental error (see Figure~\ref{fig:Dressingall}e). Furthermore, from the measurement on the 35S state another problem becomes obvious. In the reference spectrum taken in a thermal sample at $10^{12}\,\mathrm{cm}^3$, there is a small additional line red detuned by about $0.85\,\mathrm{MHz}$ to the atomic line (see Figure~\ref{fig:Dressingall}a). This line can be assigned to a molecular bound state of a Rydberg atom and a ground state atom, so called ultralong-range Rydberg molecules \cite{BBN09,BBN10}. In the BEC, this leads to loss features that are more pronounced for two reasons. First, the scaling of the Franck-Condon factor for the photoassociation of molecules with the density of ground state atoms increases the excitation probability in a BEC. And second, these bound states show a reduced lifetime at higher densities \cite{BBN11} thereby causing stronger atom losses. \\
The 44D$_{5/2},\,m_J=5/2$ state was studied at two different F\"orster defects to the resonance with the 46P$_{3/2},\,m_J=3/2$ and 42F$_{7/2},\,m_J=7/2$ pair state \cite{NBKB12,NBK12}. Since the sign of the F\"orster defect is inverted, we would expect here a dressing effect of opposite sign (see part~\ref{Dipolar}). These experiments were preformed with condensates at a four times higher atom number and therefore next to twice the peak density compared to the ones on the S-states. The better stability of the experiment at these parameters leads to a lower noise level; however, the expected effect of Rydberg dressing is even further decreased because of the higher atomic density and therefore again below the experimental noise level. In particular, we observe no significant difference between the two measurements at different F\"orster defects. Instead, for the Rydberg D-states, we observe a band of molecular states \cite{KGB14} in the spectrum (see Figure~\ref{fig:Dressingall}b). In the same range of detunings, there is a strong and several MHz broad loss feature in the measurements with the BEC, as can be seen from Figure~\ref{fig:Dressingall}d. Furthermore this loss seems to be connected to a change in aspect ratio (Figure~\ref{fig:Dressingall}f). This does not necessarily imply an asymmetric deformation of the condensate. Calculations within Thomas-Fermi approximation \cite{CD96} indicate that even an isotropic deformation of the BEC in the trap can lead to a change in aspect ratio at finite time of flight. However, it also seems unlikely that this deformation is caused by uniform losses alone, since it does not fully coincide with the observed BEC atom losses. Especially there is no such pronounced effect observed in the measurements of the Rydberg S-states (see Figure~\ref{fig:Dressingall}e) although there are equally strong atom losses present close to resonance. \\
In any case ultralong-range Rydberg molecules can possibly inhibit the observation of Rydberg dressing at red detuning for several Rydberg states. This effect becomes even more important at Rydberg states with higher principal quantum number $n$. As the binding energy is decreasing proportional to $n^{-6}$, the molecular states come closer to the atomic Rydberg state. Then polyatomic bound states \cite{BBN10,GKB14} start to play a role, eventually leading to a density dependent shift of the Rydberg line \cite{AS34,BKG13}.

\section{General challenges in realizing Rydberg dressing} \label{problems}
Several technical constraints are likely to impede an experimental observation of Rydberg dressing in our current setup. One is the unfavourable scaling of the effect with the density of ground state atoms. A low temperature sample is required to observe the small effects expected. Such samples, like Bose-Einstein condensates typically feature a high atomic density. As discussed before, the impact of Rydberg dressing is greatly reduced in strongly blockaded samples. Elements with larger background s-wave scattering length, such as cesium, allow the preparation of condensates at lower peak density \cite{WHM03}. In this case, the modification of the density distribution, however, would be even smaller due to the large mean-field interaction between the atoms. Furthermore, ultracold samples of cesium can only be prepared in optical dipole traps. Common red detuned dipole traps create differential light shifts between the ground and Rydberg state that can lead to an inhomogeneous laser detuning $\Delta$ to the Rydberg state. Therefore, a magic wavelength trap \cite{ZRS11} would be required. Instead of reducing the atomic density one possibility would be to reduce the blockade radius $r_B$ of the sample in order to tune the system just to the onset of saturation. To this end either a Rydberg state at low principal quantum number can be chosen since the $C_6$-coefficient is scaling as $n^{11}$ \cite{SSW05}. Alternatively also the excitation linewidth $\Delta f$ could in principle be increased artificially. The drawback of the first possibility is that the lifetime of the Rydberg state at the same time decreases $\propto n^3$, thereby reducing the tolerable Rydberg fraction. The latter is not practicable due to the unfavourable scaling $r_B\propto (\Delta f)^{1/6}$ and the fact that large excitation linewidths preclude realization of small laser detunings $\Delta$. Another possibility would involve a F\"orster resonance to tune the Rydberg interaction strength as discussed in part~\ref{Dipolar}. \\
Another principal problem is the long timescale required for experiments studying mechanical effects. The maximum achievable Rydberg fraction $f\approx\Omega^2/4\Delta^2$ is limited by the product of pulse length and the decay rate from the dressed state. This can be simply estimated for the experimental parameters from part~\ref{experiment}, in particular the 32S Rydberg state and a pulse length of $t=100\,\mathrm{ms}$. If we restrict the atom losses to an arbitrary value of $50\%$, we obtain the condition $f<1/(2\Gamma_rt)\approx10^{-4}$ using the decay rate $\Gamma_r=50.6\,\mathrm{kHz}$ of the 32S Rydberg state including blackbody radiation at $298\,\mathrm{K}$ \cite{RTB10}. At a detuning of $\Delta=100\,\mathrm{kHz}$, corresponding to the optimal detuning for the current density (see part~\ref{experiment}) therefore the Rabi frequency is limited to $\Omega=2\,\mathrm{kHz}$. As can be seen from Figure~\ref{fig:depOmegaDelta} the expected density change due to Rydberg dressing at these parameters is on the order of only a percent.  \\
\begin{figure}[tb]
\begin{center}
\includegraphics[width=0.6\textwidth]{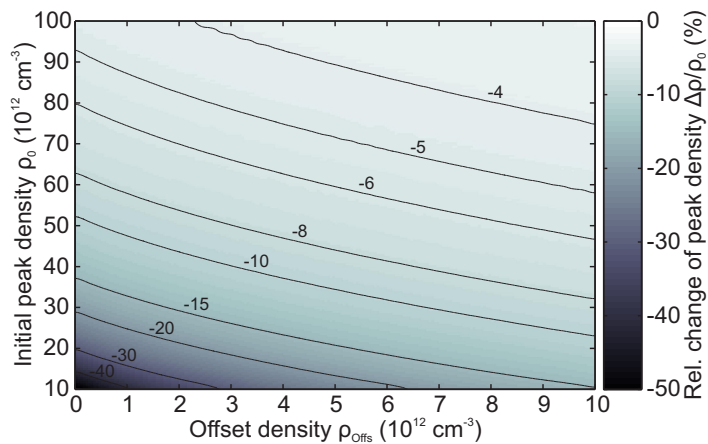}
\caption{\label{fig:deprhorhoOffs}
Dependency of Rydberg dressing of a BEC on the initial peak density $\rho_0$ and an offset density $\rho_{\mathrm{Offs}}$. The relative change of peak density $\Delta\rho/\rho_0$ in steady state is calculated for a condensate consisting of $N=2\cdot10^4$ atoms in a cylindrically symmetric trap with trap frequencies $\omega_r=2\pi\cdot 20\,\mathrm{Hz}$ and $\omega_z=2\pi\cdot 80\,\mathrm{Hz}$. The condensate is dressed detuned by $\Delta=100\,\mathrm{kHz}$ with Rabi frequency $\Omega=10\,\mathrm{kHz}$ to a repulsive Rydberg state with $C_6=6.10\cdot10^{-29}\,\mathrm{Hz}\mathrm{m}^6$, corresponding to the 32S Rydberg state.   
}
\end{center}
\end{figure}A rather technical problem finally is related to the preparation of the condensate. A very pure BEC is required since any non condensed atoms take part in the collective light shift but do not lead to a deformation of the condensate. We can account for a residual thermal cloud by simply introducing an offset density $\rho_{\mathrm{Offs}}$ in our calculations. As can be seen from Figure~\ref{fig:deprhorhoOffs} the reduction of the effect of Rydberg dressing onto the condensate is sizeable even at small offset densities $\rho_{\mathrm{Offs}}$ since the main effect according to Figure~\ref{fig:Eeff} is expected at low densities. \\
Even though at present it seems unlikely to observe Rydberg dressing at the parameters discussed above, we see two main paths towards a possible experimental realization. One would involve samples at reduced dimensionality as optical lattices \cite{MP13} thereby reducing the effective density. Adhering to three-dimensional samples instead would require a significant increase in Rabi frequency. The effect of Rydberg dressing scales roughly as $\sim\Omega^4/\Delta^3$ (equations~\ref{gdress} and \ref{eq:Edresslim}) while the decay rate of the dressed state is only scaling as the Rydberg fraction $\sim\Omega^2/\Delta^2$. By increasing the laser detuning $\Delta$ by the same amount as the Rabi frequency $\Omega$ one can therefore expect a stronger impact at the same loss rate. For the parameters discussed above, a modification of the peak density on the order of ten percent would therefore require a more than hundred times larger laser power.
   
\section{Conclusion}
In summary, we have presented a simple model of a Rydberg dressed system. This model provides an easy qualitative and quantitative description of the collective effects arising. We studied the impact of Rydberg dressing onto a three-dimensional axially symmetric Bose-Einstein condensate, a situation that is experimentally relevant, depending on different parameters. Our model allows to intuitively understand the dependency of the maximum effect on various experimental parameters. The limit of Thomas-Fermi approximation in the case of Rydberg dressing was identified by a comparison with a full numerical solution of the Gross-Pitaevskii equation. We proposed the extension to Rydberg states close to a F\"orster resonance. Here further work is required in order to understand possible collective effects arising in the F\"orster coupling. Finally we described our current experimental situation in view of a possible realization of Rydberg dressing. Principle as well as practical challenges were demonstrated and, as a conclusion, possible solutions were discussed. 


\ack
We thank H. P. B\"uchler and his group for fruitful discussions and support. This work is funded by the Deutsche Forschungsgemeinschaft (DFG) within the SFB/TRR21 and the project PF~381/4-2. We also acknowledge support by the ERC under contract number 267100 and A.G. acknowledges support from E.U. Marie Curie program ITN-Coherence 265031.

\section*{References}

\bibliographystyle{unsrt}


\end{document}